\documentclass[12pt]{iopart}

\usepackage[dvips]{graphicx}

\usepackage{iopams}

\begin{document}

Presented at 5-th International Workshop on Science with the new generation high energy gamma-ray experiments, Frascati, Italy, June 2007

\title[]{Searching for Kaluza-Klein Dark Matter Signatures in the LAT Electron Flux}

\author{A.A.~Moiseev$\S$,
E.A.~Baltz$^*$,
J.F.~Ormes$\P$,
and L.G.~Titarchuk$^+$}

\address{$\S$ CRESST and AstroParticle Physics Laboratory, NASA/GSFC, Greenbelt, USA}
\address{$^*$ KIPAC, Menlo Park, USA }
\address{$\P$ University of Denver, Denver, USA}
\address{$^+$ NRL, Washington DC, USA}
\address{Corresponding author: alexander.a.moiseev@nasa.gov}

\begin{abstract}
We present here the prospects for the GLAST Large Area Telescope (LAT) detection of the signature of the lightest Kaluza-Klein particle (LKP).
It decays by direct annihilation into electron-positron pairs that may be detectable in the high energy electron flux.
We discuss the LAT capability for detecting the high energy (20 GeV - ~1 TeV) cosmic ray electron flux and we analyze the LAT sensitivity to detect LKP-produced electrons for various particle
masses. We include an analysis of the diffusive propagation of the electrons in the galaxy.
\end{abstract}

\section{Introduction.}
The nature of dark matter and dark energy is one of the most exciting and critical problems in modern astrophysics. A number of theoretical models
 predict the existence of dark matter in different forms; so the experimental detection will be crucial. A large number of
 experiments are ongoing and planned  to detect dark matter, directly and indirectly, both at accelerators and in  space, where they search for dark matter signatures in cosmic radiation (see \cite{Baltz} and references therein).
In this paper we explore the capability of GLAST Large Area Telescope (LAT), scheduled for launch in the beginning of 2008,
to detect the signature of dark matter in the high energy cosmic ray electron flux. We have previously demonstrated that LAT will be a  powerful detector of cosmic ray electrons,
and will provide measurement of their flux with high statistical confidence \cite{moiseevICRC}. We should mention that LAT is not designed to distinguish electrons from positrons, so we refer to their  sum as electrons for simplicity. LAT will detect  $\approx 10^{7} $ electrons per year above 20 GeV with the energy resolution 5-20\%.
Such good statistics permits detection of spectral features, among which could be ones caused by  exotic sources such as Kaluza-Klein particles which manifest higher spatial dimensions. The possibility of the annihilation of the Lightest Kaluza-Klein Particles (LKP), which can be a stable and viable dark matter candidate, directly into electron-positron pairs, was investigated in, e.g. \cite{Baltz} and \cite{Cheng}. They estimated that electron-positron pairs are produced in approximately 20\% of the annihilation cases, which makes the observations viable within the model assumptions.

There are some indications of spectral features in the electron spectrum observed by ATIC \cite{ATIC} and PPB-BETS \cite{BETS} around 300-500 GeV, as well as in the positron spectrum measured by HEAT \cite{HEAT}, encouraging us for  measurements with the LAT.

\section{LAT Capability to detect cosmic ray electrons}
It was demosnstrated earlier that the LAT can efficiently detect cosmic ray electrons \cite{moiseevICRC}. Being a gamma-ray telescope, it intrinsically is an electron spectrometer. The main problem is to separate the electrons from all other species, mainly protons. In order to keep the hadron contamination in the detected electron flux under 10\%, the hadron-electron separation power must be $ 10^{3}-10^{4} $. At very high energy (above a few TeV) the diffuse gamma-radiation could be a potential background, but it will be effectively eliminated by the LAT AntiCoincidence Detector. LAT's onboard trigger accepts all events with the detected energy above $ \approx  20$ GeV, which is very important in order to have unbiased data sample. We took advantage of this LAT feature and explored the instrument sensitivity in the energy range above 20 GeV. It is also good to mention that there should be no problem with albedo and geomagnetic variation in this energy range.

We have developed a set of analysis cuts that select electrons and applied them to simulated LAT data. The approach was based on using the difference in the shower development between hadron-initiated and electron-initiated events. In order to obtain the instrument response function for electrons we  simulated the electron spectrum incident on LAT and applied our selections. In the energy range from 20 GeV to 1 TeV  the effective geometric factor (for electrons) after applying our cuts is $0.2 - 2$ m$^{2}$
sr  and energy resolution is 5-20\% depending on the energy. We also applied the selections to the simulated cosmic ray flux (LAT background flux is used, see \cite{Ormes}) and determined the residual hadron contamination to be $ \approx3\% $ of the remaining electron flux.

In order to test the approach, we run an independent simulation of the incident flux and used the response function obtained  to reconstruct the spectrum. For the simulation of the electron flux we used the diffusion equation solution given in \cite{Atoyan}. Fig.1 shows our spectrum reconstruction for the simulated electron flux collected during 1 year of LAT observations. The flux originated from an "hypothetical" single burst-like source, $ 2\times10^{5} $ years old, at a distance of 100 pc. The diffusion coefficient D was assumed to be energy dependent as $ D=D_{0}\left( 1+E/E_{0}\right) ^{0.6} $ with $ D_{0}=10^{28}$ cm$^{2}$s$^{-1}$. The expected spectral cutoff for this model is $ \approx1.2 $ TeV. With the demonstrated precision in the spectrum reconstruction we should be able to recognize the specific features which can be associated with dark matter.

\begin{figure}[here]
\begin{center}
\includegraphics[scale=0.35]{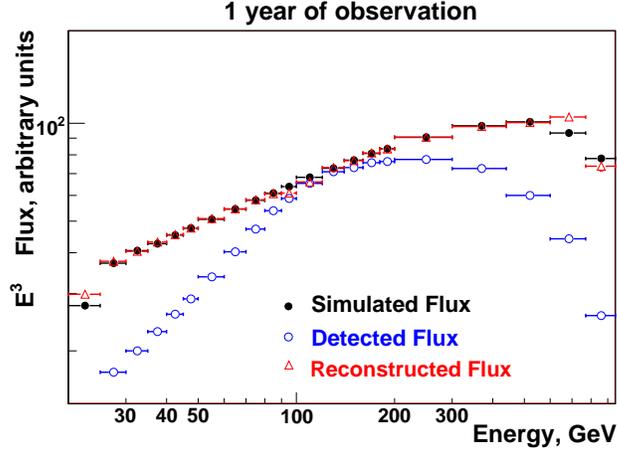}
\caption{Simulated electron flux reconstruction for LAT}
\end{center}
\end{figure}

\section{Diffusive propagation of  the LKP signal}
Now we want to include the contribution from the LKP annihilation into the electron flux, and see how it can be detected by the LAT. We consider this contribution as a continuous injection from a single point source of monoenergetic electrons, assuming dark matter clumpiness, and determine the effect of their propagation through  space to the Earth. After that we will test LAT response to such a spectrum, varying the source parameters LKP mass and distance.
We treat the propagation of electrons using the stationary diffusion equation which in the spherical symmetric case
is presented as
\begin{equation}
 \frac{D}{r^2}\frac{\partial}{\partial r}r^2 \frac{\partial f}{\partial r} +
\frac{\partial}{\partial \gamma}(Pf) =Q
\label{main_eq}
\end{equation}
where $f(r,\gamma)$ is  a distribution  of electron number over
$\gamma=E/m_ec^2$ and radius $r$ at time $t$.

We assume that the continuous energy loss is dictated by synchrotron and inverse Compton losess
\begin{equation}
-d\gamma/dt=P(\gamma)=p_2\gamma^2
\label{p_eq}
\end{equation}
where
\begin{equation}
p_2=5.2\times 10^{-20}\frac{w_0}{1{\rm eV/cm^3}}~{\rm s}^{-1}
\label{p2_eq}
\end{equation}
and $\omega_0 \simeq 1$ eV/cm$^3$ (see \cite{Atoyan} for the propagation details)

We choose the energy-dependent diffusion coefficient in the form
\begin{equation}
D(\gamma)= D_0(1+\gamma/\gamma_{g})^{\eta}~{\rm cm}^{2}{\rm s}^{-1}
\label{D_eq}
\end{equation}
where $\gamma_{g}$ and
 $D_0$ are model parameters.

We derive the general solution of Eq. ({\ref{main_eq}) for arbitrary source function  in the factorized form
\begin{equation}
Q=\varphi(r)\psi(\gamma)
\label{Q_fact}
\end{equation}
 as well as a solution for a $\delta-$ function injection, i.e. for
 \begin{equation}
 Q(r,\gamma)=\delta(\gamma-\gamma_{\ast})\delta(r-r_0)/4\pi r_0^2.
\label{Q_delta}
\end{equation}

Using Eq. (\ref{main_eq}) it  can be shown that the Green's function $G_{0,\gamma_{\ast}} (r,\gamma)$,
as a  solution for the delta-function source (see  Eq. \ref{Q_delta}) is presented as
\begin{equation}
G_{0,\gamma_{\ast}} (r,\gamma)= \frac{1}{D_0}\frac{R_{0}[r,u(\gamma, \gamma_{\ast})]}{\gamma^2}
\label{Green}
\end{equation}
where $R_{0}[r,u(\gamma, \gamma_{\ast})]$ and
$u(\gamma, \gamma_{\ast})$  are determined by  Eq. (\ref{R_0}) and  Eq. (\ref{sol_char}) respectively:
\begin{equation}
R_{0}(r,u)=\frac{1}{8u(\pi u)^{1/2}}\exp(-r^2/4u) .
\label{R_0}
\end{equation}

\begin{equation}
u(\gamma, \gamma_0)=\frac{D_0}{p_2}\int_\gamma^{\gamma_0}\frac {(1+\gamma/\gamma_{g})^\eta d\gamma}{\gamma^2}.
\label{sol_char}
\end{equation}
Integral (\ref{sol_char}) can be presented in the analytical form in two cases:
for $\gamma_{g} \to \infty$ or $\eta=0$  it is (see \cite{ST80})
\begin{equation}
u(\gamma, \gamma_0)=\frac{D_0}{p_2}\left(\frac{1}{\gamma}-\frac{1}{\gamma_{0}}\right).
\label{sol_char_inf}
\end{equation}
and for $\eta=0.5$ it is
$$
u(\gamma, \gamma_0)=\frac{D_0}{p_2}\times
$$
\begin{equation}
\left[\frac{1}{\gamma}
\left(1+\frac{\gamma}{\gamma_{g}}\right)^{1/2}-
\frac{1}{\gamma_0}\left(1+\frac{\gamma_0}{\gamma_{g}}\right)^{1/2} +
\frac{1}{\gamma_{g}}\ln\frac{(\gamma_g/\gamma)^{1/2}+(1+\gamma_g/\gamma)^{1/2}}
{(\gamma_g/\gamma_0)^{1/2}+(1+\gamma_g/\gamma_0)^{1/2}}\right]
\label{sol_char_05}
\end{equation}

\subparagraph{}

\begin{figure}[here]
\begin{center}
\includegraphics[scale=0.4]{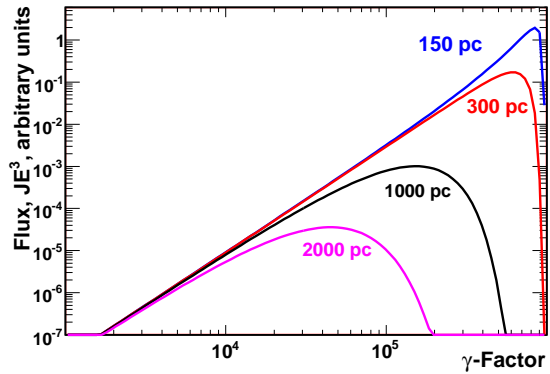}
\caption{Propagation of the signal from LKP with the mass of 500 GeV, from different distances}
\end{center}
\end{figure}

Fig.2 illustrates the  solution obtained and shows the propagation of the signal from annihilation of LKP, with mass of 500 GeV, for different distances. Due to losses during propagation, both the peak energy and the signal magnitude decrease with the increasing distance.  There will be a superposition of contributions from dark matter clumps at different distances, which could reveal themselves as bumps in the spectrum, but the closest clump should determine the edge in the spectrum which should be clearly seen.

\section{Prospects for LAT to observe LKP}

Our analysis of the LAT capability for detection of electrons demonstrated the low residual hadron contamination in the resulting electron flux ($ <3\% $, see Section 2). The contamination from gamma-radiation will also be negligible, so the dominant background in the search for LKP signature will consist of only "conventional" electron flux. Now we can apply the LAT capability for electron detection  to one of the dark matter models, using the scenario given in \cite{Baltz} as an example. The LKP annihilation will be seen as a line (edge) in the electron spectrum of magnitude proportional to $ m_{LKP}^{-6} $\\
\begin{equation}
 \frac{dN_{e}}{dE_{e}}=\frac{Q_{line}( m_{LKP}) }{b(E_{e})}\theta(m_{LKP}-E_{e})
\end{equation}

 \begin{equation}
 \sim\langle\sigma v\rangle\left( \frac{\rho_{0}}{m_{LKP}}\right) ^{2}\left( \frac{1}{E_{e}^{2}}\right) \theta(m_{LKP}-E_{e})\sim m_{LKP}^{-6}
\end{equation}
where $\langle \sigma v\rangle$ is the total annihilation cross section of LKP, and $Q_{line}$ is the rate of electron and positron injection from direct LKP annihilation.

Using the numbers from \cite{Baltz}: boost factor $B=5$ and $\rho_{local}=0.4$ GeV cm$^{3}$, the magnitude of the signal after propagation is estimated as
\begin{equation}
\left( \frac{dN_{e}}{dE_{e}}\right) \approx \frac{9.5\times10^{8}}{m_{LKP}^{6}\left[ {\rm GeV}\right] }m^{-2}{\rm s}^{-1}{\rm sr}^{-1}{\rm GeV}^{-1}.
\end{equation}

\begin{figure}[here]
\begin{center}
\includegraphics[scale=0.42]{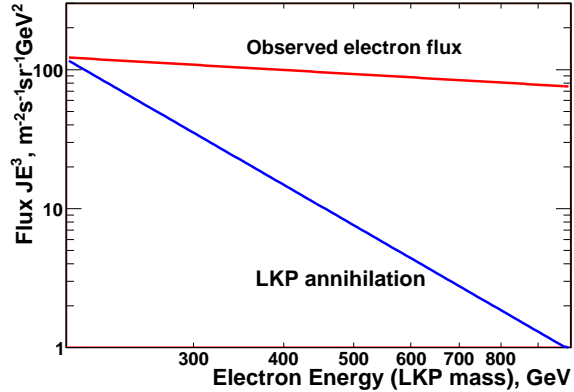}
 \caption {Expected electron flux from LKP annihilation, along with the observed electron flux}
\end{center}
\end{figure}

\begin{figure}[here]
\begin{center}
\includegraphics[scale=0.6]{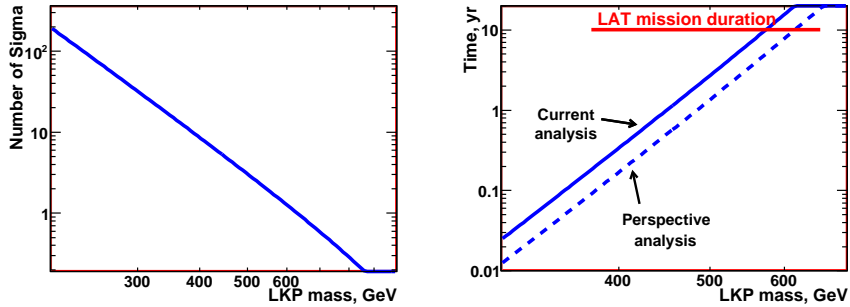}
\caption{LKP detection in LAT electron spectrum. Left panel - detection significance for 1 year. Right panel - time needed to detect LKP feature with $5 \sigma$ confidence}
\end{center}
\end{figure}

Fig.3 shows the expected signal from LKP vs. the $m_{LKP}$ plotted along with the "conventional" electron flux for the comparison. Of course, even in this optimistic model, the background dominates over the signal, but we now determine what will be the LAT sensitivity.  Fig.4 shows the significance of LKP detection in the LAT-detected electron flux in one year of ovservation, and the observation time needed to detect LKP feature with $5\sigma$ confidence for a source at 100 pc. We can conclude that 600 GeV is probably the heaviest LKP which could be observed within the constrains assumed. Taking into account that for thermal freeze-out, LKP mass in the range 600-700 GeV is preferred, the feasible window for LKP mass in the LAT search is rather narrow.

Now we illustrate our analysis by adding the signal from LKP  (mass 300 GeV), from a single clump at a distantce of 100 pc, to the "conventional" electron flux shown in Fig.1, using the solution obtained in Section 3. The result is shown in Fig.6, with a clear signature of presence of a monoenergetic component. This is a very favorable situation, but to some extent consistent with references \cite{ATIC} and \cite{BETS}.

\begin{figure}[here]
\begin{center}
\includegraphics[scale=0.4]{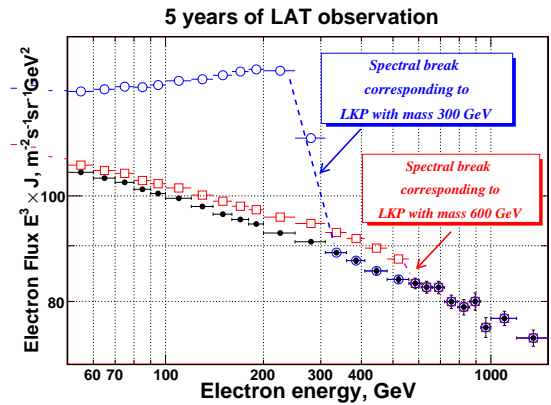}
\caption{Illustration of the simulated reconstructed LAT electron spectrum with the presence of signal from LKP with the mass of 300 GeV and 600 GeV, for five years of observations. Black filled circles - "conventional" electron flux, blue open circles - with added signal from 300 GeV LKP, red oped squares - with added signal from 600 GeV LKP }
\end{center}
\end{figure}

\section{Conclusion}
We analyzed the capability of the LAT to detect high energy cosmic ray electrons and applied it to the model of LKP direct annihilation into electron-positron pairs. Using the estimates for this model as given in \cite{Baltz} as an example to demonstrate the detection feasibility, we estimated that within this model, the LAT will be able to recognize the LKP-caused spectral edge in the electron spectrum up to a LKP mass of 600-700 GeV. The results obtained  can be applied to any dark matter model where  electrons are produced in order to estimate the LAT sensitivity. The important feature is that the dominating background in these measurements will be
only the "conventional" electron flux.

We want to thank all LAT team members, and especially the members of LAT Dark Matter Science Working Group for their support and valuable suggestions. We are grateful to Robert Hartman and Jan Conrad for their comments on this paper.


\section {References}
\begin{thebibliography}{99}

\bibitem{Baltz} E.A. Baltz and D. Hooper, JCAP, 7(2005),1

\bibitem{moiseevICRC} A.A. Moiseev, J.F. Ormes and I.V. Moskalenko, Proceedings of 30-th ICRC,
Merida, Mexico, 2007

\bibitem{Cheng} H.-C. Cheng, J.L. Feng, and K.T. Matchev, Phys. Rev. Letters, 89, 21 (2002)

\bibitem{ATIC} J. Chang et al., Proceedings of 29-th ICRC, 3,1, Pune, India, 2005

\bibitem{BETS} K. Yoshida et al., Proceedings of 30-th ICRC, Merida, Mexico, 2007

\bibitem{HEAT} S.W. Barwick et al., Astrophys. J. 498 (1998), 779

\bibitem{Ormes} J.F. Ormes et al., The First GLAST Symposium, Edotors S. Rita, P. Michelson, and C. Meegan, AIP 921 (2007), 560

\bibitem{Atoyan} A.M. Atoyan, F.A. Aharonian, and H.J. Volk, Physical Review D, 52(1995), 3265

\bibitem{ST80} R.A. Sunyaev and L.G. Titarchuk, Astronomy and Astrophysics, 86 (1980), 121

\end{thebibliography}
\end{document}